\documentstyle[prb,epsfig,multicol,aps,amsmath,array]{revtex}
\renewcommand{\narrowtext}{\begin{multicols}{2} \global\columnwidth20.5pc}
\renewcommand{\widetext}{\end{multicols} \global\columnwidth42.5pc}
\begin{document}

\title{Wigner molecules in polygonal quantum dots: A density functional study}

\author{E. R\"as\"anen, H.~Saarikoski, M.~J.~Puska, and R.~M.~Nieminen}

\address{Laboratory of Physics, Helsinki University of Technology,
P.O. Box 1100, FIN-02015 HUT, FINLAND} 

\date{\today}
\maketitle

\begin{abstract}
We investigate the properties of many-electron systems in two-dimensional polygonal 
(triangle, square, pentagon, hexagon) potential wells by using the 
density functional theory. The development of the ground state electronic structure 
as a function of the dot size is of particular interest.
First we show that in the case of two electrons, the Wigner molecule formation 
agrees with the previous exact diagonalization studies.
Then we present in detail how the spin symmetry breaks in polygonal geometries as the
spin density functional theory is applied. In several cases with more than two
electrons, we find 
a transition to the crystallized state, yielding coincidence with the number of 
density maxima and the electron number. We show that this transition density, which 
agrees reasonably well with previous estimations, is rather insensitive to both the 
shape of the dot and the electron number.
\end{abstract}

\vspace{0.8cm}

\narrowtext

\section{Introduction}
The research of nanoscale electronic structures has been expanding
continuously. Quantum dots \cite{qd} represent basic electron systems
that have been fabricated using semiconductor materials for almost fifteen years. 
Because the confinement of electrons in quantum dots or 'artificial atoms'
can be varied at will, they have became a playground in which the basic 
physics of interacting electrons can be surveyed and theoretical models
can be tested.

In quantum dots the correlation effects between electrons have to be considered
carefully because the external confinement is remarkably weaker than
in real atoms, where the independent electron model with mean-field theories usually gives
good results. As the confinement strength is lowered, the mutual Coulomb interaction
becomes gradually dominant and at a certain point, the electron density begins to
exhibit localization to classical positions in order to minimize the 
interaction. This phenomenon corresponds to the Wigner crystallization (WC) in a 
two-dimensional electron gas (2DEG) (Ref. 2).
According to quantum Monte Carlo simulations, the crystallization occurs
when the 2D electron density $n$ has decreased such that the density parameter
$r_s > 37$. Here  $n=1/(\pi r_s^2)$ and $r_s$ is given in units of the effective Bohr radius
$a^*_{B}=\hbar^2\epsilon/m^*{e^2}$, where $\epsilon$ is the dielectric
constant and $m^*$ is the effective electron mass, specific to the
semiconductor material in question.
In 2DEG with impurities the broken translational invariance has been
shown to result in the WC at a much larger electron density with $r_s \simeq 7.5$
(Ref. 3). In quantum dots the transition to the WC has been
predicted to occur at even higher electron densities \cite{poly1,egger,yanno,reimann,reusch}.
One of the questions in this context is how the shape 
and the electron number of a two-dimensional quantum dot affect the crystallization.

The weak-confinement limit in quantum dots has been studied with various theoretical
methods, including exact diagonalization \cite{poly1,mikhailov}, quantum Monte Carlo
(QMC) \cite{egger,VMC}, and unrestricted Hartree-Fock (UHF) \cite{yanno,reusch} techniques, as
well as the spin density functional theory (SDFT) \cite{reimann,oma,koskinen,akbar}. 
In this regime the SDFT allows the formation of spin density waves (SDW), 
{\em i.e.} the breaking of the spin symmetry \cite{koskinen}, leading to a lower total
energy in the system.
In our earlier work for a parabolic six-electron quantum dot, we examined the 
energy difference between the polarized and paramagnetic spin states, and showed that
the SDW solution agrees well with the QMC results, in contrast to the symmetry-preserved
DFT solution \cite{oma}.
However, the problem in the SDFT calculations is the fact that only the z-component of 
the total spin can be specified. Therefore, one may ask if a mixed state of several 
eigenstates, corresponding to different $S$ with a fixed $S_z$, is physically meaningful, 
an argument presented by Hirose and Wingreen \cite{hirose}. 
In fact, a mixed-symmetry state is not an eigenstate of the Hamiltonian, but the lowest 
state of a well-defined mixture of symmetries is a functional of the density 
\emph{at the time of preparation of the state} \cite{barth}.
The symmetry-broken electron structure thus gives more accurate approximations for the energy
and describes the internal nature of the many-body wave function better than the 
symmetry-preserved solution \cite{mottel}.

The criterion for the WC in quantum dots may be determined with several attributes. 
Egger {\em et al.} \cite{egger} considered three criteria yielding similar results
in their QMC analysis for parabolic quantum dots. 
They observed the electron density in real space and
searched for the confinement at which the shell structure began to form. 
In addition, they monitored a quantity depending on the pair-correlation function,
and changes in the energy spectra. Localization may also be
observed directly by examining probability densities of single electrons \cite{VMC}.
Creffield {\em et al.} \cite{poly1} have studied the systems of two electrons 
confined by polygonal 2D infinite-barrier wells, and their
criterion for the onset of the WC is the appearence of a local density minimum at 
the center of the dot. In the (S)DFT calculations the criterion should be based on 
the density (spin densities).
However, it became evident in our study that the criterion by Creffield {\em et al.} 
can not be applied for polygonal dots containing more than two electrons, because the 
electron density has maxima at the corners of the dot also at very high electron densities.  

Quantum dots are usually modeled by restricting a certain
number of electrons to a 2D plane and assuming the confining potential to have
a parabolic shape. 
An example of more general modeling is the above-mentioned exact diagonalization
study by Creffield {\em et al.} \cite{poly1}. Moreover, Akbar and Lee \cite{akbar} 
used the SDFT to study square quantum dots which have a small finite extent in the
third dimension perpendicular to the square. In the case of two electrons, 
they found a good agreement with the results by Creffield {\em et al.}
For two- and four-electron dots, Akbar and Lee estimated the onset of the WC to
occur at $r_s \simeq 6$.

In the present work we employ the SDFT to investigate the properties of two-dimensional 
quantum dots with a general polygonal confinement. We concentrate on the WC
in the weak-confinement limit, which is obtained simply by increasing 
the side length of the dot. In the numerical calculations we apply a recently 
developed real-space approach \cite{mika}.
As a symmetry-unrestricted method it is flexible regarding the applied geometry
and allows also SDW solutions.
In the regime of the spin symmetry-broken solutions, we find that further weakening of
the confinement leads to electron densities with
as many maxima as there are electrons in the system. We show that the appearance
of this behavior can be used, at least in several cases, consistently 
as a criterion for the onset of the WC for quantum dots of various shapes and 
different electron numbers.

The outline of the paper is as follows. In Sec. \ref{sec2} we present the theoretical model
and the computational method of our calculations. From the results we first compare
the DFT, {\em i.e.} spin-compensated calculations for a two-electron dot to the exact 
diagonalization results.
Then we employ the SDFT and present the symmetry-broken solutions first
for a two-electron dot, and thereafter for larger systems.
A summary and discussion are given in Sec \ref{sec4}.

\section{Methods} \label{sec2}

The quantum dot material is chosen to be GaAs with the effective electron mass $m^*=0.067m_e$ 
and the dielectric constant $\epsilon=12.4$. The effective Bohr radius $a^*_{B}$ is thus 
9.79 nm. The Hamiltonian of a many-electron system in a polygonal potential well is written as
\begin{equation}
H=\sum^N_{i=1}\left[-\frac{\hbar^2}{2m^*}\nabla^2_i+V_{\rm ext}({\mathbf r}_i)\right]
+\sum^N_{i<j}\frac{e^2}{\epsilon|{\mathbf r}_i-{\mathbf r}_j|},
\label{ham}
\end{equation}
where the external potential has a simple form
\begin{equation}
V_{\rm ext}(x,y)=\left\{\begin{gathered}
0, \text{ in the dot}\\
\infty, \text{ elsewhere}.
\end{gathered}
\right.
\end{equation}
The effective mass approximation (EMA) used with the Hamiltonian (\ref{ham})
represents an alternative to the constant-interaction model \cite{kouwenhoven}, in which 
the Coulomb interaction between the electrons is assumed to be independent of the electron 
number $N$. The EMA has been shown to be a reliable approximation if the confinement
is not particularly strong \cite{france}.

In the SDFT formalism, the electron density is solved self-consistently with the Kohn-Sham
equations \cite{dft1,dft2}. To approximate the exchange-correlation energy functional, we use
the local spin density approximation (LSDA) with the interpolation form by Tanatar and Ceperley 
\cite{tanatar} for the 2DEG. Within the EMA, the single-particle Schr\"odinger equation of
the Kohn-Sham scheme reads as
\begin{equation}
\Bigg[-\frac{\hbar^2}{2m^*}\nabla^2+
V^{\sigma}_{\rm eff}({\mathbf r})\Bigg]\psi_{i,\sigma}({\mathbf r})=
\epsilon_i\psi_{i,\sigma}(\mathbf{r}),
\label{schr}
\end{equation}
where $V^{\sigma}_{\rm eff}$ is the effective potential for spin $\sigma$ containing
the external potential and the Hartree and exchange-correlation potentials of the 
electron-electron interactions. In the spin-compensated calculations
(equal spin densities), the scheme reduces to the standard density functional
theory (DFT) within the local density approximation (LDA).

In the self-consistent Kohn-Sham scheme, we perform calculations in real space by using 
two-dimensional point grids. The differential equations are discretized with finite
differences \cite{beck}, and the procedure is efficiently accelerated with 
multigrid techniques \cite{brandt} to solve the Poisson and the single-particle 
Schr\"odinger equations. Applying the multigrid scheme in the latter case
is a fairly complicated task because both the eigenfunctions and
the eigenvalues have to be solved simultaneously. In order to avoid nonlinearity problems, 
the Rayleigh Quotient Multigrid (RQMG) method \cite{mandel} is used for the solution of
the eigenpair corresponding to the lowest eigenvalue. We employ this method with
a recent generalization to an arbitrary number of lowest eigenenergy states \cite{mika}.
The discretized eigenvalue equation is solved by
minimizing the Rayleigh quotient $\big<\psi|H|\psi\big>/\big<\psi|\psi\big>$ 
on the {\em finest} grid, using the coarser grids to remove the lower frequency components 
of the error. The technique reduces remarkably the number of relaxation sweeps needed
for solving the Schr\"odinger equation. Other advantages of the real-space solver 
are its flexibility with the boundary conditions and good suitability for parallel 
computing. 

\section{Wigner crystallization of two electrons} \label{sec3}

\begin{figure}
\includegraphics{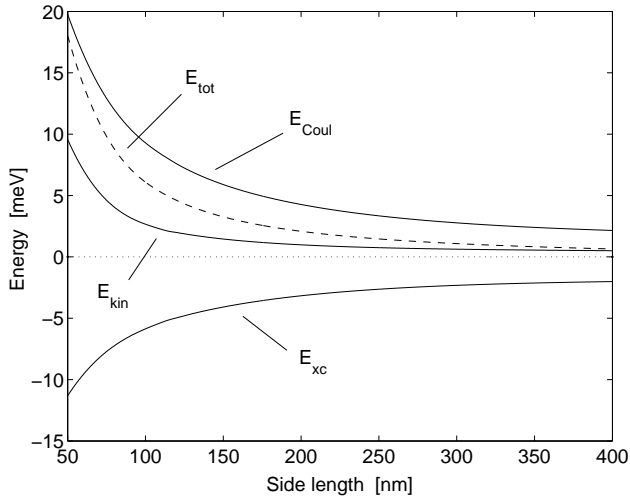}
\vspace{0.05cm}
\caption{Energy composition in a square two-electron quantum dot as a function of the dot size.}
\label{energy}
\end{figure}

First we perform DFT calculations on two-electron polygonal quantum dots by
setting the spin densities equal ($n_{\uparrow}=n_{\downarrow}$) to prevent the
breaking of the spin symmetry.
The ensuing total energy with its decomposition
to Coulomb, kinetic and exchange-correlation energies are given in Fig. \ref{energy}
for a square dot. As predicted, the Coulomb energy becomes increasingly more dominant over 
the kinetic energy as the side length $L$ of the dot is enlarged and the WC is expected
to occur.

\begin{figure}
\includegraphics[width=17cm]{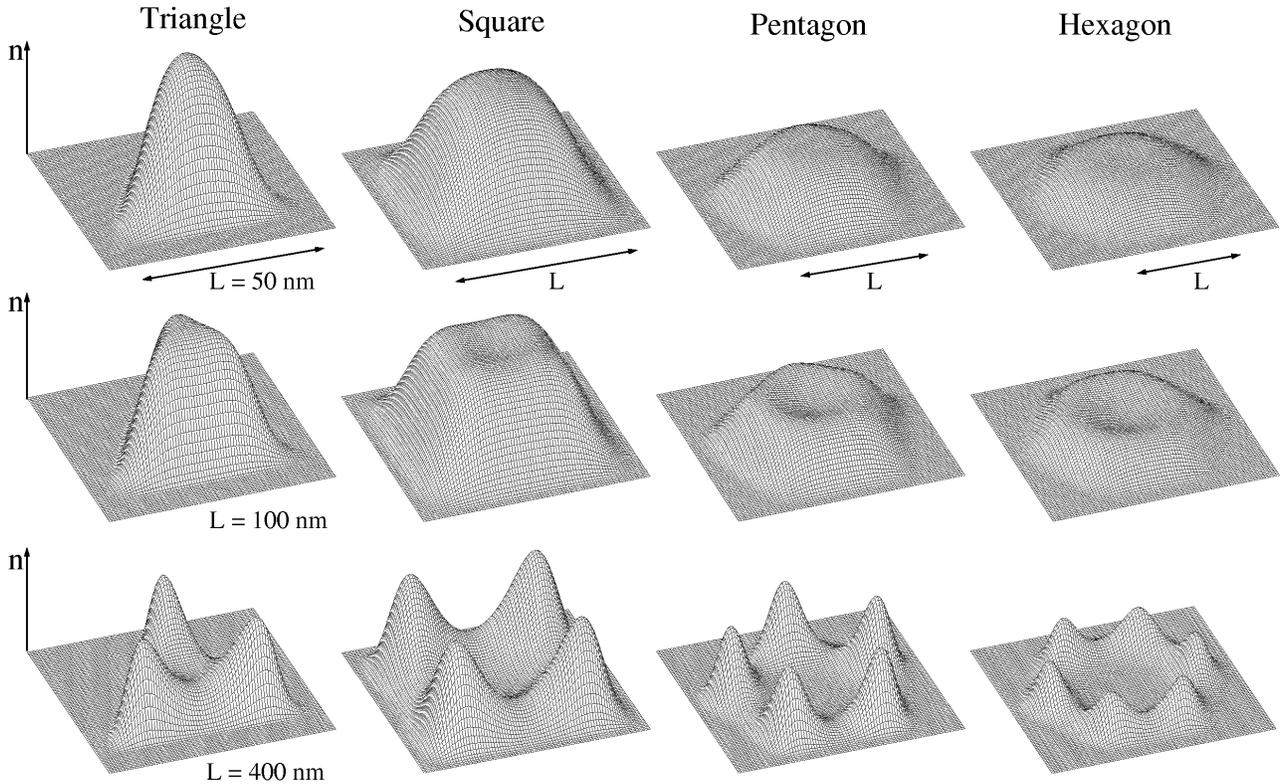}
\vspace{0.3cm}
\caption{Electron densities in polygonal two-electron quantum dots with different sizes. In the
square, pentagon, and hexagon the amplitudes have been multiplied by a factor 2.}
\vspace{1cm}
\label{densities}
\end{figure}

The ground state electron density distributions for the triangular, square, pentagonal, and 
hexagonal dots at three side lengths $L$ = 50, 100, and 400 nm are presented in Fig. \ref{densities}.
In the small dots the electron density is lumped at the center, whereas the large dots represent
Wigner-molecule-like behavior, the density being localized near the corners in order
to minimize the dominating Coulomb interaction. The localization is seen to depend strongly on 
the area of the dot. Creffield {\em et al.} \cite{poly1} defined the system to be a Wigner 
molecule, when a local minimum first appears at the center. According to our calculations,
this occurs, for instance, in the square as the side length is about 100 nm ($\simeq 10 a^*_{B}$),
which agrees with the exact diagonalization results. This qualitative consistency establishes 
the applicability of the density functional approach to small systems considered in this study.

We define the density parameter as  $r_s=\sqrt{A/N\pi}$, where $A$ is the
area of the polygon. In the case of $n$ corners and a side length $L$ we thus get
$r_s=\frac{L}{2}\sqrt{\frac{n}{N\pi}\cot{\frac{\pi}{n}}}$. By applying the criterion presented
by Creffield 
{\em et al.} \cite{poly1} for the WC transition point, we find the value of $r_s\sim{3}$ 
in all four geometries. Akbar and Lee \cite{akbar} employed the SDFT for square 
2D quantum dots with an additional harmonic confinement along the z-axis. They used a
more rigorous criterion for the WC, {\em i.e.} the breaking of
connections between the density maxima, and estimated
the critical value of $r_s\sim{6}$ for the transition point. In spite of the very
different estimation technique, this result is in a qualitative agreement with our
$r_s\sim{3}$.

Intuitively, the localization of two electrons into all the corners of 
a polygonal potential well might first appear as a slightly odd result.
Jefferson and H\"ausler 
\cite{jefferson} have explained the phenomenon with effective charge-spin models.
They suggested that the low-energy manifold of a system of
strongly correlated electrons can be described properly with an extended single-band Hubbard
model. For example, in a square two-electron dot the tV-Hamiltonian transforms into
the following effective Hamiltonian,
\begin{equation}
H_{\rm eff}=\tilde{E}_0+\left(\Delta{e^{i2\Phi}}R_{\pi/2}+\textrm{h.c.}\right),
\end{equation}
where $R_{\pi/2}$ rotates the electrons at opposite corners on a diagonal by $\pi/2$.
The electron pair may thus tunnel between the ground state configurations with an amplitude
modulated by a factor $e^{i2\Phi}$. This explains the four-peak structure of the electron
density in the Wigner limit, predicted already by Bryant \cite{bryant}.
Diagonalization of $H_{\rm eff}$ gives a good approximation for the ground 
state energies obtained from the tV-Hamiltonian \cite{creffield2}.

\widetext

\narrowtext

\section{Symmetry-broken solutions}

\subsection{Two-electron dot}

Next we perform the same calculations as above but without the restriction 
$n_{\uparrow}=n_{\downarrow}$, and consider still the ground state solution, 
for which $S_z=0$. Comparison of the new total energies with the 
spin-compensated results as a function of the dot size reveals an interesting transition 
to a lower energy state. 
At this point, representing already a Wigner-crystallized distribution, 
the spin symmetry breaks and the result is a SDW-like ground state. 

\begin{figure}
\includegraphics[width=8cm]{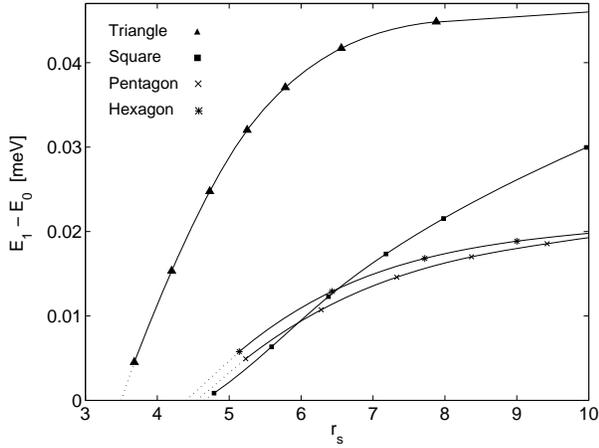}
%\vspace{0.2cm}
\caption{Total energy differences between the DFT ($E_1$) and SDFT ($E_0$)
solutions in polygonal two-electron quantum dots of four geometries.}
\label{poly}
\end{figure}

The relative energy differences between the spin symmetric and
SDW-like solutions, corresponding to our DFT and SDFT calculations respectively,
are shown in Fig. \ref{poly} for all the considered geometries. In the triangular well the
transition to the symmetry-broken ground state occurs at a remarkably smaller size than 
in the other three geometries. More precisely, for the triangle we get the transition 
at $r_s\simeq{3.5}$ and for the square, pentagon, and hexagon at $r_s\sim{4.5}$. 

\begin{figure}
\includegraphics[width=8cm]{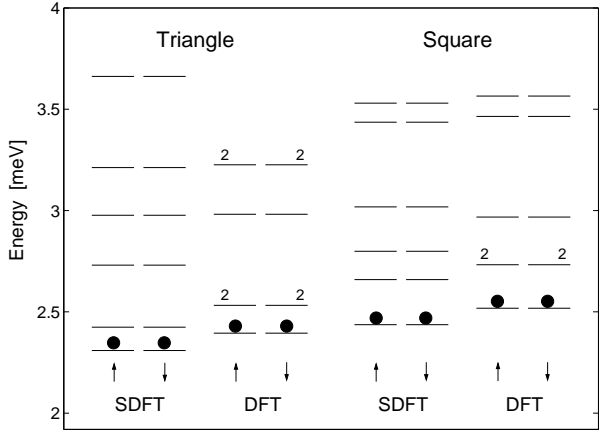}
\vspace{0.2cm}
\caption{Lowest energy levels of triangular and square two-electron quantum dots 
at $r_s\sim 8$.}
\label{lev}
\end{figure}

\begin{figure}
\includegraphics[width=8cm]{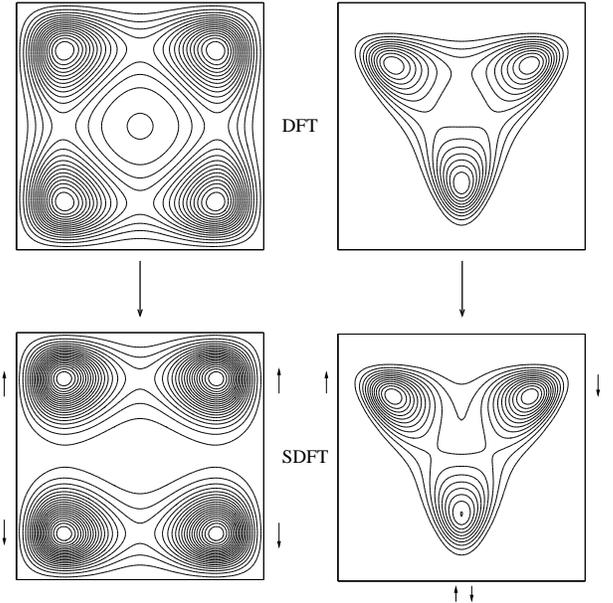}
\vspace{0.2cm}
\caption{Difference in the electron densities between the DFT (up) and SDFT (down) 
solutions in a square and triangular two-electron quantum dot at $L=$ 400 nm.
The spin alignments are shown in the SDFT case.}
\label{sdw}
\end{figure}

In order to explain this behavior, one may first examine the lowest 
one-electron energy states, shown in Fig. \ref{lev}
for the triangular and 
square quantum dots in the symmetry-broken SDFT ground state, as well as in the
symmetry-preserved DFT solution. 
In the latter state, the three-fold geometry produces more low-lying degenerate
levels in the triangle than the four-fold geometry in the square.
In the SDFT calculations these degeneracies are split
such that the energy levels become pronouncedly spread in 
the triangular geometry, whereby the lowest levels
are pushed more efficiently downwards in the triangle than in the square.
There is also a qualitative difference between the symmetry-broken electron densities 
in these geometries. As shown in Fig. \ref{sdw}, the spin-up and spin-down 
densities are totally separated in the square, whereas in the triangle they share
a corner. In the triangular geometry, the breaking of the spin symmetry can thus lower 
the energy via the exchange-correlation and Coulomb contributions relatively 
more, and with a relatively smaller cost in the kinetic energy than in the square.
Nevertheless, in none of these geometries the breaking of the spin 
symmetry enlarges the Fermi gap, contrary to the SDW formation in large, 
parabolic quantum dots \cite{koskinen}.

\begin{figure}
\includegraphics[width=8cm]{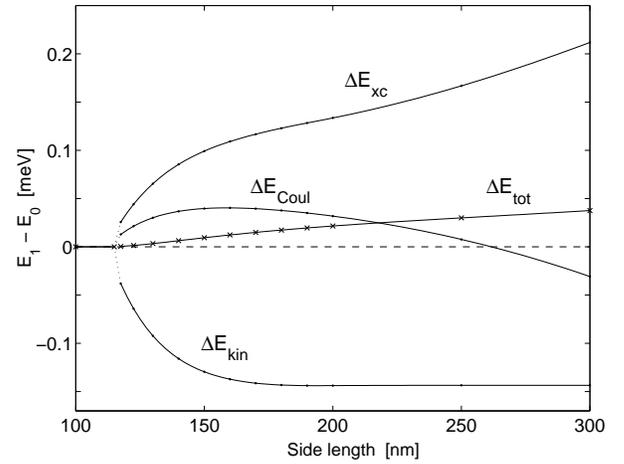}
\vspace{0.1cm}
\caption{Composition of the energy difference between the DFT ($E_1$) and
SDFT ($E_0$) solutions of a square
two-electron quantum dot as a function of the dot size.}
\label{poly4}
\end{figure}

The composition of the energy difference between the symmetry-preserved and the
symmetry-broken states are presented in Fig. \ref{poly4} for a square dot.
Naturally, the change in the exchange-correlation energy favors and the change
in the kinetic energy opposes the transition. The behavior of the Coulomb 
energy is interesting: its strong decrease actually initiates the 
breaking of the spin symmetry.
However, as the dot is made larger than $L\sim{250}$ nm, the Coulomb energy is higher
in the SDW-like than in the symmetry-preserved state. The phenomenon can be
understood by having a further look at the electron density distributions in the square
(Fig. \ref{sdw}). In the SDFT solution, the electron density is shifted from the region
between the opposite spin directions towards the corners. At small dot sizes this
decreases the Coulomb repulsion between the charge peaks in the adjacent corners 
more than the repulsion increases inside the peaks. At large
distances the opposite is true.

\vspace{0.3cm}

\newpage

\subsection{$N>2$}

Then we consider some special cases with more than two electrons.
The next geometry-independent magic configuration after $N=2$ is a six-electron dot.
It represents an interesting point of comparison with the results obtained for 
a parabolic quantum dot in the weak-confinement limit. 
We find that the spin symmetry is broken at $r_s\simeq$ 3.8, 3.1, 4.6, 4.9 
in the triangle, square, pentagon, and hexagon quantum dots, respectively. 

In the parabolic dot with $V_{\rm ext}(r)={1\over 2}\omega_0^2r^2$, the $r_s$ parameter 
can be estimated from $\omega_0^2=e^2/(e\pi\epsilon_0\epsilon{m^*}r_s^3\sqrt{N})$
(Ref. 12).
In pursuance of our earlier work for this quantum dot \cite{oma}, 
the SDW formation was not found until $r_s\simeq{6.6}$.
The sharp corners in the confinement seem thereby favor the transition to the
symmetry-broken state. In the six-electron case, however, the triangular geometry is more stable 
against the transition than the square one. A square with $N=6$ represents an inconvenient
combination, similar to the triangle with $N=2$, in which the electrons cannot be evenly 
divided to the corners of the polygon. As the number of the corners increases further,
the transition shifts to higher $r_s$ values, approaching the point of the SDW formation 
in the parabolic quantum dot with a circular symmetry.

\begin{figure}
\includegraphics[width=8cm]{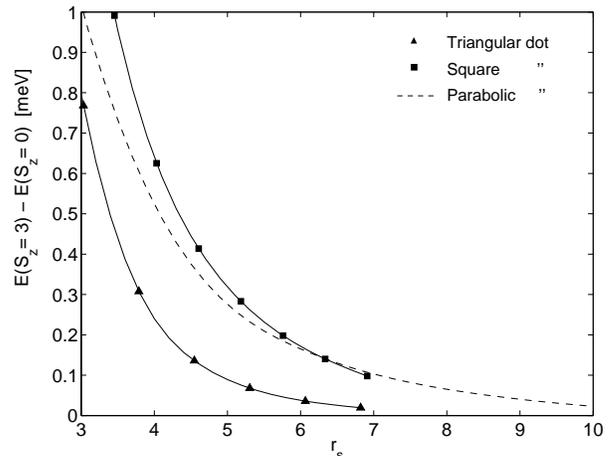}
%\vspace{0.2cm}
\caption{Total energy differences between the $S_z=3$ and $S_z=0$
states in the triangular (triangle markers), square (square markers), and 
parabolic (dashed line) six-electron quantum dots.}
\label{polar}
\end{figure}

For $N=6$, we consider also the possibility of spin-polarization, {\em i.e.} the
$S_z=3$ state becoming the ground state in the low density limit. The energy differences
between the polarized ($S_z=3$) and paramagnetic ($S_z=0$) states for triangular and
square geometries as a function of $r_s$ are shown in Fig. \ref{polar}.
For comparison, the SDFT results for the parabolic quantum dot \cite{oma} are 
also presented, the latter showing spin-polarization at $r_s > {12}$.
We were not able to obtain well-converged results for the triangle and square
quantum dots at large $r_s$ values. Therefore we can only
speculate by extrapolation that the polarization could occur in the triangle and square 
slightly earlier than in the parabolic quantum dot.

\begin{figure}
\includegraphics[width=8cm]{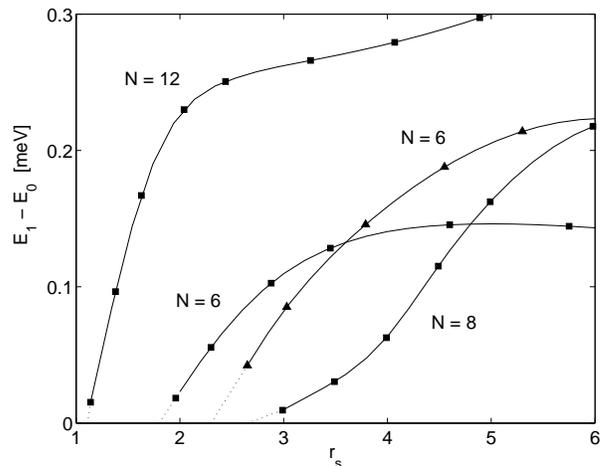}
%\vspace{0.2cm}
\caption{Total energy differences between the DFT ($E_1$) and SDFT ($E_0$)
solutions in square quantum dots with $N=$ 6, 8, and 12 (square markers).
Results for the $N=6$ triangle quantum dot are also given (triangle markers).}
\label{many1}
\end{figure}

\begin{figure}
\includegraphics[width=8cm]{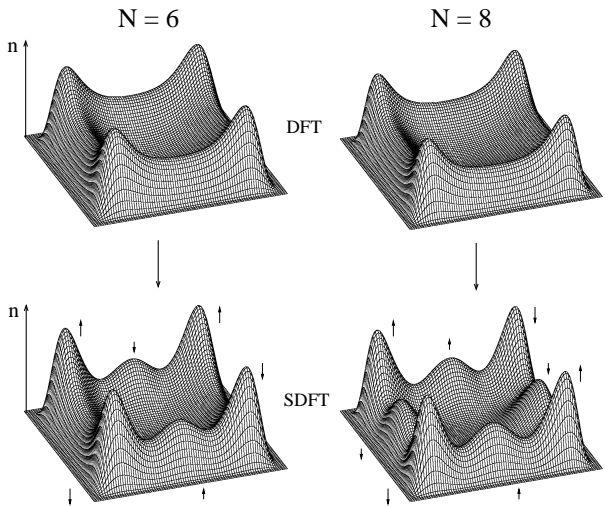}
\vspace{0.3cm}
\caption{Electron densities of the DFT (up) and SDFT (down) 
solutions in the $N=6$ and $N=8$ square quantum dots with side lengths $L=$ 300 nm.
The spin alignments are shown in the SDFT case.}
\label{many2}
\end{figure}

Besides the geometry, we can study how the number of electrons affects the breaking of
spin symmetry. 
First we consider a square dot with $N=$ 6, 8, and 12, 
which all correspond to completely filled shells. 
Fig. \ref{many1} shows the energy difference between
the spin symmetry-preserved and -broken solutions as a function of the $r_s$. 
For $N=6$ and $N=12$,
the spin symmetry breaks at $r_s\simeq 1.7$ and $r_s\simeq 1.1$, respectively,
whereas the ground state of the $N=8$ dot remains spin symmetric until 
$r_s\simeq 2.8$. However, the energy difference grows rapidly in this dot,
being considerably larger than in the $N=6$ dot at $r_s\simeq 10$.
In the large dots the SDFT solutions show pronounced localization of the 
spin densities as can be seen in Fig. \ref{many2}. For $N=$ 6 and 8, 
the number of maxima in the total electron densities equals to the number
of electrons, leading to $\pi$ and $\pi/2$ rotational symmetries in
these systems, respectively. The spin symmetry can be considered to be
broken more completely in the $N=8$ dot, where the density peaks
with the same spin are located on diagonally opposite vertices, 
in contrast to the $N=6$ dot where they lie on adjacent corners.
The interaction is thus minimized more efficiently in the $N=8$
dot, corresponding to a relatively
rapid decrease of the total energy shown in Fig. \ref{many1}.
For comparison, the results for a triangular quantum dot
with $N=6$ are also presented. 
In that system, the
increase in the energy difference resembles the behavior of the $N=8$ square dot,
reflecting a similar symmetry-broken geometry (see Fig. \ref{maxima} below).

\vspace{0.2cm}

\newpage

\vspace{0.1cm}

After the breaking of the spin symmetry, there can be seen only four density 
maxima in the corners of the $N=6$ and $N=8$ square quantum dots, 
resembling the DFT solution 
(the upper row of Fig. \ref{many2}). The dot size has to be increased 
substantially before the maxima in the middle of the edges appear 
(the lower row). We can observe the same behavior in 
the $N=6$ triangle and $N=10$ pentagon, in both which the spin symmetry
breaks at $r_s\simeq 2.3$.
Their density distributions
at large $r_s$ values are given in Fig. \ref{maxima}. In all these four cases,
the number of density maxima equals to the number of electrons in the system.
Therefore the appearance of the last density peaks can be considered as
the final stage in the onset of the WC in the SDFT formalism. 

\begin{figure}
\includegraphics[width=8cm]{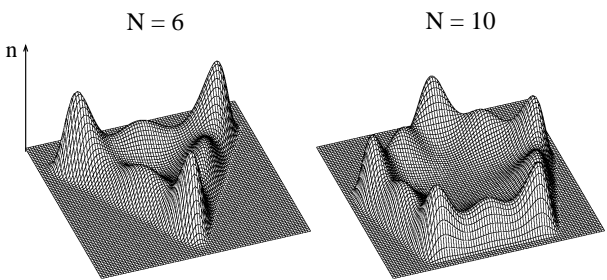}
\vspace{0.2cm}
\caption{Electron densities at $r_s\sim{8}$ in triangular and pentagonal quantum dots with $N=6$ and $N=10$,
respectively.}
\label{maxima}
\end{figure}

\vspace{0.1cm}

\begin{figure}
\includegraphics[width=8cm]{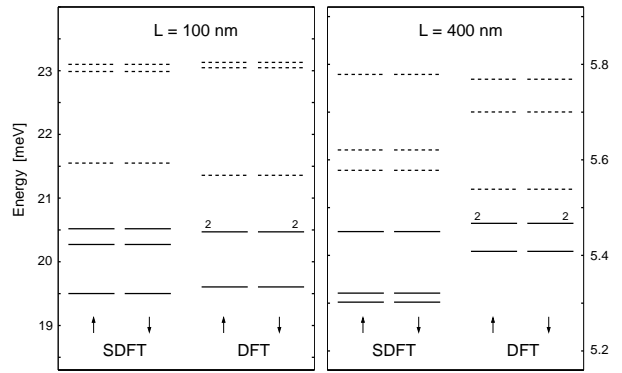}
\vspace{0.25cm}
\caption{Lowest single-particle energy levels in a $N=6$ square quantum dot with
side lengths $L=$ 100 and 400 nm. Solid and dashed lines correspond to the 
occupied and unoccupied staes, respectively. The levels are nondegenerate, except
the doubly degenerate levels denoted by the numbers (2).}
\label{many3}
\end{figure}

In order to analyze the appearance of the last density peaks, we show in Fig. \ref{many3}
the lowest one-electron energy levels for the $N=6$ square quantum dot with  
side lengths $L=$ 100 and 400 nm, corresponding to $r_s\sim$ 2 and 9, respectively. 
At the smaller size, the spin symmetry has already broken and
split the DFT degeneracies. As the dot is made larger, the two lowest states become closer to
each other and are remarkably lowered in comparison with the symmetry-preserved DFT solution.
This condensation occurs similarly in all the dots, in which the electron density localizes
to a number of maxima coinciding with the number of electrons. The appearance
of the last density maxima thus drives the lowest energy levels towards degeneracy. The complete 
degeneracy would be the ultimate state for the Wigner crystal. In Fig.  \ref{many3} 
one can also notice that the Fermi gap is considerably larger in the symmetry-broken solution 
than in the symmetry-preserved case, resembling the situation in large, parabolic quantum dots 
\cite{koskinen}.

\vspace{0.1cm}

\vspace{0.1cm}

We have carefully determined the $r_s$ values at which the last maxima appear
and found astonishingly similar values for the different systems studied,
although the breaking of the spin symmetry occurs on a broad $r_s$ scale.
The critical values of $r_s\simeq$ are 3.8 and
4.0 for $N=6$ and $N=8$ square dots, respectively, and $r_s\simeq$ 3.9
for both the $N=6$ triangle and the $N=10$ pentagon.
In the case of two-electron dots, the above criterion for the WC cannot be applied, but
the onset of the spin symmetry-broken state gives an reasonable estimate of  
$r_s\simeq$ 3.5 for the triangular $N=2$ dot and $r_s\simeq$ 4.5 for the
other $N=2$ polygonal quantum dots. Our estimate of $r_s\simeq$ 4.0
for the WC transition point is consistent with the results for 
small, parabolic quantum dots \cite{egger,yanno,reimann,reusch,mikhailov,VMC,oma,koskinen}.
It is also clearly smaller than $r_s\simeq{7.5}$ obtained for the 
fluid-solid transition in 2DEG containing impurities \cite{chui}.

\section{Summary} \label{sec4}
 
We have studied the electronic properties of polygonal two-dimensional quantum dots
by employing the spin density functional theory. The numerical calculations are 
performed with a symmetry-unrestricted real-space scheme. Especially, we have been focused
on the behavior of these systems at the weak-confinement limit, where the role
of the electron-electron interactions becomes dominating and eventually leads
to the formation of the so-called Wigner molecules.
 
First we have shown that the density functional theory is capable to reproduce,
in agreement with the exact diagonalization studies,
the behavior of the electron density in polygonal two-electron quantum dots
as the spatial size of the potential well increases.
 
The spin density functional theory leads
inevitably to the breaking of the spin symmetry. For different geometries and
different electron numbers, this occurs in a wide range of average electron
densities or $r_s$ parameters. The spin symmetry-broken density shows for
certain geometries and electron numbers a gradual transition, such that the
number of density maxima coincides with the number of electrons. We use
the the appearance of the last density maxima as the criterion for the
Wigner crystallization and obtain $r_s\simeq 4.0$ for the critical density. 
This value does not depend strongly on the geometry, nor the electron number of 
the quantum dot, and is in agreement with Quantum Monte Carlo results.

\begin{acknowledgments}
This research has been supported by the Academy of Finland through its Centers of
Excellence program (2000-2005).
\end{acknowledgments}

\widetext \end{document}